\documentclass[english,onecolumn]{IEEEtran}
\usepackage[T1]{fontenc}
\usepackage[latin9]{inputenc}
\usepackage{amsmath}
\usepackage{amsthm}
\usepackage{amssymb}
\usepackage{esint}

\makeatletter
\theoremstyle{plain}
\newtheorem{thm}{\protect\theoremname}
\theoremstyle{plain}
\newtheorem{lem}[thm]{\protect\lemmaname}
\theoremstyle{plain}
\newtheorem{cor}[thm]{\protect\corollaryname}

\usepackage{mathrsfs}
\usepackage{algorithm,algpseudocode}

\usepackage{colortbl}
\definecolor{lightgray}{rgb}{0.9,0.9,0.9}
\definecolor{lightred}{rgb}{1,0.8,0.8}
\definecolor{lightgreen}{rgb}{0.6,1,0.6}
\definecolor{lightyellow}{rgb}{1,1,0.5}
\definecolor{lightgrey}{rgb}{0.8,0.8,0.8}

\allowdisplaybreaks[1]

\makeatother

\usepackage{babel}
\providecommand{\corollaryname}{Corollary}
\providecommand{\lemmaname}{Lemma}
\providecommand{\theoremname}{Theorem}

\begin{document}
\title{Pointwise Redundancy in One-Shot Lossy Compression via Poisson Functional
Representation}
\author{Cheuk Ting Li\thanks{This work was partially supported by an ECS
grant from the Research Grants Council of the Hong Kong Special Administrative
Region, China {[}Project No.: CUHK 24205621{]}.\newline C. T. Li
is with the Department of Information Engineering, The Chinese University
of Hong Kong, Hong Kong SAR of China (e-mail: ctli@ie.cuhk.edu.hk).}}
\maketitle
\begin{abstract}
We study different notions of pointwise redundancy in variable-length
lossy source coding. We present a construction of one-shot variable-length
lossy source coding schemes using the Poisson functional representation,
and give bounds on its pointwise redundancy for various definitions
of pointwise redundancy. This allows us to describe the distribution
of the encoding length in a precise manner. We also generalize the
result to the one-shot lossy Gray-Wyner system.
\end{abstract}

\medskip{}

\section{Introduction}

Lossy source coding concerns the problem of compressing a source such
that the reconstruction is close to the source with respect to a distortion
measure. For fixed-length codes where the compression has a fixed
number of bits, the optimal compression rate in the asymptotic setting
where the blocklength tends to infinity is given by the rate-distortion
function \cite{shannon1959coding,cover2006elements}, whereas the
optimal second-order term is given in terms of the $d$-tilted information
\cite{ingber2011dispersion,kostina2012fixed}.

Variable-length codes, where the length of the compression can depend
on the source, have also been studied. Variable-length codes are natural
in universal source coding settings \cite{davisson1973universal,ziv1977universal},
where the source sequence $X_{1},\ldots,X_{n}$ follows an unknown
distribution, and the encoding length should adapt to the distribution.
Variable-length codes are also useful in non-universal settings where
the source distribution is known. For example, for lossy compression
with a fixed upper-bound on the excess distortion probability, \cite{kostina2015variable}
studied a construction where we assign empty codewords in case of
excess distortion, resulting in a reduction of the expected length. 

For another example, $D$-semifaithful codes \cite{ornstein1990universal,yu1993rate}
concerns the setting where the distortion must be bounded almost surely.
Such an almost sure distortion constraint necessitates the use of
variable-length codes, since fixed-length codes generally cannot achieve
the rate-distortion function. In \cite{zhang1997redundancy}, it was
shown that the (average) rate redundancy $\ln n/n+o(\ln n/n)$ can
be achieved for $D$-semifaithful codes. For a pointwise converse
result, it has been shown in \cite{kieffer1991sample} that $\mathrm{liminf}_{n}\ell_{n}(X^{n})/n\ge R(D)$
almost surely, where $\ell_{n}(X^{n})$ denotes the encoding length
of the source sequence $X^{n}$, and $R(D)$ is the rate-distortion
function. Pointwise redundancy of $D$-semifaithful codes in the finite
blocklength setting has been studied in \cite{kontoyiannis2000pointwise}
(see Section \ref{sec:redundancy}). The convergence rate of the encoding
length to the rate-distortion function has been characterized in \cite{dembo2001critical}.
The case for sources with memory was investigated in \cite{ouguz2012pointwise}.
Also refer to \cite{silva2021universal,mahmood2023lossy} for results
on universal $D$-semifaithful codes. Variable-length codes are also
useful for entropy-constrained quantization \cite{gish1968asymptotically,chou1989entropy}.

Another motivation for variable-length codes is that there is a logarithmic
gap between the expected length of one-shot lossy compression (there
is only one source symbol $X$) under the expected distortion constraint
$\mathbb{E}[d(X,Y)]\le D$ \cite{sfrl_trans}, and the optimal asymptotic
rate given by the rate-distortion function $R(D)$. More precisely,
it was shown in \cite{sfrl_trans}, via a construction called \emph{Poisson
functional representation}, that there exists a prefix-free code with
expected length
\begin{equation}
\le R(D)+\log(R(D)+1)+6.\label{eq:sfrl_rd}
\end{equation}
Also see \cite{harsha2010communication,braverman2014public} for related
channel simulation results, and \cite{posner1971epsilon} for a related
result on epsilon entropy. This is a lossy analogue of variable-length
lossless compression, where Huffman coding \cite{huffman1952method}
gives a constant gap between the expected length of one-shot compression
and the optimal asymptotic rate. This would not be possible for fixed-length
codes, which has an optimal one-shot length arbitrarily larger than
the optimal asymptotic rate even for lossless compression.

In this paper, we utilize the Poisson functional representation \cite{sfrl_trans,li2021unified}
to construct one-shot variable-length lossy source coding schemes.
We study three notions of pointwise redundancy, namely \emph{pointwise
rate redundancy} (PRR) measuring the difference between the encoding
length and $R(D)$ (studied in \cite{kontoyiannis2000pointwise,ouguz2012pointwise}),
\emph{pointwise source-wise redundancy} (PSR) measuring the difference
between the encoding length and the  $d$-tilted information in the
source \cite{kontoyiannis2000pointwise,kostina2012fixed,csiszar1974extremum}
(studied in \cite{kontoyiannis2000pointwise}), and the \emph{pointwise
source-distortion-wise redundancy} (PSDR) measuring the difference
between the encoding length and a measure of the amount of information
needed to encode the source within a given distortion. We give bounds
on these pointwise redundancies, where the PSDR admits an especially
simple bound. We also generalize the results to the one-shot lossy
Gray-Wyner system \cite{gray1974source,sfrl_trans}.

\medskip{}

\subsection*{Notations}

Entropy is in bits, and logarithm is to the base $2$. Write $[a]_{+}:=\max\{a,0\}$.
Write $\{0,1\}^{*}:=\bigcup_{k=0}^{\infty}\{0,1\}^{k}$ for the set
of sequences with any length. For $c\in\{0,1\}^{*}$, write $|c|$
for its length. For random variables $X,Y$, write $P_{X}$ for the
distribution of $X$, and $\iota_{X;Y}(x;y)=\log\frac{\mathrm{d}P_{X,Y}}{\mathrm{d}P_{X}\times P_{Y}}(x,y)$
for the information density. Write $X^{n}=(X_{1},\ldots,X_{n})$.

\medskip{}

\section{Main Result}

A \emph{lossy compression scheme with positive integer description}
for the source $X\in\mathcal{X}$, $X\sim P_{X}$ with reconstruction
space $\mathcal{Y}$ consists of a (possibly stochastic) encoder given
as a conditional distribution $P_{K|X}$ from $\mathcal{X}$ to $\mathbb{Z}_{>0}$,
and a decoding function $g:\mathbb{Z}_{>0}\to\mathcal{Y}$. Given
the source $X$, the encoder produces the description $K\in\mathbb{Z}_{>0}$
using $P_{K|X}$, and sends it to the decoder which reconstructs $\tilde{Y}=g(K)$.
We often impose an \emph{expected distortion constraint}. Let $d:\mathcal{X}\times\mathcal{Y}\to[0,\infty)$
be a distortion function. Then we may require that $\mathbb{E}[d(X,\tilde{Y})]\le D$
for a fixed $D\in\mathbb{R}$. Note that if we want to bound the probability
of excess distortion $\mathbb{P}(d(X,\tilde{Y})>D)\le\epsilon$ instead,
then we can still consider it as an expected distortion constraint
by introducing a new distortion measure $d'(x,y)=\mathbf{1}\{d(x,y)>D\}$,
and imposing the condition $\mathbb{E}[d'(X,\tilde{Y})]\le\epsilon$.

For a \emph{lossy compression scheme with variable-length description},
we instead have the stochastic encoder $P_{M|X}$ from $\mathcal{X}$
to $\{0,1\}^{*}$ (now the description is $M\in\{0,1\}^{*}$ instead
of $K$), and a decoding function $g:\{0,1\}^{*}\to\mathcal{Y}$.
We can choose whether to impose the prefix-free condition or not.
If we impose the prefix-free condition, then it is required that $M\in\mathcal{C}$,
where $\mathcal{C}\subseteq\{0,1\}^{*}$ is a prefix-free codebook.
Note that a scheme with positive integer description can be converted
into a variable-length scheme without the prefix-free condition \cite{szpankowski2011minimum},
since $K$ can be encoded into $\lfloor\log K\rfloor$ bits given
by the binary representation of $K$ without the leading digit. It
can also be converged into a prefix-free variable-length scheme, for
example, by using the Elias delta code \cite{elias1975universal}
that encodes $K$ into $\le\log K+2\log(\log K+1)+1$ bits.

We now present the main result in this paper.

\medskip{}

\begin{thm}
\label{thm:main}Fix any $P_{X}$, $P_{Y|X}$ and $Q_{Y}$ satisfying
$P_{Y|X}(\cdot|x)\ll Q_{Y}$ for $P_{X}$-almost all $x$'s. Consider
any finite collection of functions $\psi_{i}:\mathcal{X}\times\mathcal{Y}\times\mathbb{Z}_{>0}\to\mathbb{R}$
that are nondecreasing in the third argument (i.e., $\psi_{i}(x,y,k)$
is nondecreasing in $k$ for any fixed $x,y$) for $i=1,\ldots,\ell$.
Then there exists a lossy compression scheme with positive integer
description $K\in\mathbb{Z}_{>0}$ and reconstruction $\tilde{Y}$
such that
\begin{equation}
\mathbb{E}\big[\psi_{i}(X,\tilde{Y},K)\big]\le\mathbb{E}\big[\psi_{i}(X,Y,\ell J)\big]\label{eq:main_exp}
\end{equation}
for $i=1,\ldots,\ell$, where $(X,Y)\sim P_{X}P_{Y|X}$, and $J\in\mathbb{Z}_{>0}$
is distributed as
\begin{align}
 & J\,|\,\{X=x,Y=y\}\nonumber \\
 & \sim\mathrm{Geom}\bigg(\bigg(\frac{\mathrm{d}P_{Y|X}(\cdot|x)}{\mathrm{d}Q_{Y}}(y)+1\bigg)^{-1}\bigg).\label{eq:main_geom}
\end{align}
\end{thm}
\medskip{}

Theorem \ref{thm:main} is versatile. For example, if we want to impose
an expected distortion constraint, we take $\psi_{i}(x,y,k)=d(x,y)$.
If we want to impose an excess distortion probability constraint,
we take $\psi_{i}(x,y,k)=\mathbf{1}\{d(x,y)>D\}$. If we use a fixed-length
code with $n$ bits, and want to bound the probability that $K$ cannot
be encoded into $n$ bits, we take $\psi_{i}(x,y,k)=\mathbf{1}\{k>2^{n}\}$.
If we use a variable-length code without prefix-free condition, and
want to bound the expected length, we take $\psi_{i}(x,y,k)=\log k$.
If we want the prefix-free condition, we may instead take $\psi_{i}(x,y,k)=\log k+2\log(\log k+1)+1$
(using Elias delta code \cite{elias1975universal}). We will later
see in Section \ref{sec:redundancy} how we can choose $\psi_{i}$
in order to bound the pointwise redundancy.

We now describe the construction of the coding scheme for Theorem
\ref{thm:main}, which utilizes the \emph{Poisson functional representation}
\cite{sfrl_trans,li2021unified}. Here we use a construction mostly
similar to \cite[Theorem 2]{sfrl_trans}, with a refined analysis
using techniques in \cite{li2021unified}. Let $0\le T_{1}\le T_{2}\le\cdots$
be a Poisson process with rate $1$ (i.e., $T_{1},T_{2}-T_{1},T_{3}-T_{2}\stackrel{iid}{\sim}\mathrm{Exp}(1)$),
and $\bar{Y}_{1},\bar{Y}_{2},\ldots\stackrel{iid}{\sim}Q_{Y}$ be
independent of $(T_{i})_{i}$. The process $(\bar{Y}_{i},T_{i})_{i}$
(which is a marked Poisson process) serves as the ``random codebook''
of the coding scheme. Given $X$, the encoder outputs 
\begin{equation}
K:=\mathrm{argmin}_{i}\frac{T_{i}}{(\mathrm{d}P_{Y|X}(\cdot|X)/\mathrm{d}Q_{Y})(\bar{Y}_{i})}.\label{eq:k_scheme}
\end{equation}
Given $K$, the decoder outputs $Y=\bar{Y}_{K}$.

The following result was given in \cite[Equation (29)]{li2021unified}
(after substituting $j=1$).

\medskip{}

\begin{lem}
[Poisson functional representation \cite{li2021unified}]\label{lem:pfr}Consider
two distributions $P\ll Q$. Let $0\le T_{1}\le T_{2}\le\cdots$ be
a Poisson process with rate $1$, and $\bar{U}_{1},\bar{U}_{2},\ldots\stackrel{iid}{\sim}Q$
be independent of $(T_{i})_{i}$. Let $U=\bar{U}_{K}$, where
\[
K:=\mathrm{argmin}_{i}\frac{T_{i}}{(\mathrm{d}P/\mathrm{d}Q)(\bar{U}_{i})}.
\]
We treat $T_{i}/0=\infty$ here. Then $U\sim P$, and $K$ has the
following conditional distribution given $U$:
\[
K|\{U=u\}\sim\mathrm{Geom}\bigg(\Big(\mathbb{E}\Big[\max\Big\{\frac{\mathrm{d}P}{\mathrm{d}Q}(u),\,\frac{\mathrm{d}P}{\mathrm{d}Q}(U')\Big\}\Big]\Big)^{-1}\bigg),
\]
where $U'\sim Q$.
\end{lem}
\medskip{}

Lemma \ref{lem:pfr} shows that $Y|X\sim P_{Y|X}$, and
\begin{align}
 & K\,|\,\{X=x,\,Y=y\}\nonumber \\
 & \sim\!\mathrm{Geom}\bigg(\!\!\Big(\mathbb{E}\Big[\!\max\!\Big\{\frac{\mathrm{d}P_{Y|X}(\cdot|x)}{\mathrm{d}Q_{Y}}(y),\frac{\mathrm{d}P_{Y|X}(\cdot|x)}{\mathrm{d}Q_{Y}}(Y')\!\Big\}\!\Big]\Big)^{\!-1}\!\bigg),\label{eq:k_geom}
\end{align}
where $Y'\sim Q_{Y}$. We have
\begin{align*}
 & \mathbb{E}\Big[\max\Big\{\frac{\mathrm{d}P_{Y|X}(\cdot|x)}{\mathrm{d}Q_{Y}}(y),\,\frac{\mathrm{d}P_{Y|X}(\cdot|x)}{\mathrm{d}Q_{Y}}(Y')\Big\}\Big]\\
 & \le\mathbb{E}\Big[\frac{\mathrm{d}P_{Y|X}(\cdot|x)}{\mathrm{d}Q_{Y}}(y)+\frac{\mathrm{d}P_{Y|X}(\cdot|x)}{\mathrm{d}Q_{Y}}(Y')\Big]\\
 & =\frac{\mathrm{d}P_{Y|X}(\cdot|x)}{\mathrm{d}Q_{Y}}(y)+1.
\end{align*}
Therefore, the distribution $P_{K|X=x,Y=y}$ in \eqref{eq:k_geom}
is first order stochastically dominated by $P_{J|X=x,Y=y}$ in \eqref{eq:main_geom}.

Since the encoder and decoder cannot share common randomness, they
cannot agree on a random codebook $\mathfrak{P}:=(\bar{Y}_{i},T_{i})_{i}$.
Therefore, we have to ``derandomize'' the scheme and fix a codebook.
By invoking Carath{\'{e}}odory's theorem in a manner similar to \cite[Theorem 2]{sfrl_trans},\footnote{The ordinary Carath{\'{e}}odory's theorem would require $\ell+1$
points. Here we require one fewer point since we only need inequality
instead of equality.} there exist fixed choices of codebooks $\mathfrak{p}_{1},\ldots,\mathfrak{p}_{\ell}$
and $w_{1},\ldots,w_{\ell}\ge0$ with $\sum_{j}w_{j}=1$ such that
\[
\mathbb{E}[\psi_{i}(X,Y,\ell K)]\ge\sum_{j=1}^{\ell}w_{j}\mathbb{E}[\psi_{i}(X,Y,\ell K)\,|\,\mathfrak{P}=\mathfrak{p}_{j}].
\]
The encoder then randomize among these codebooks, by choosing the
$j$-th codebook $\mathfrak{p}_{j}$ with probability $w_{j}$. To
allow the decoder to know which codebook is used, the encoder transmits
$K'=\ell(K-1)+j$ instead of $K$, where $j$ is the index of the
codebook chosen. We have $K'\le\ell K$, and $\mathbb{E}[\psi_{i}(X,Y,\ell K)]\le\mathbb{E}[\psi_{i}(X,Y,\ell J)]$
since $P_{K|X,Y}$ is first order stochastically dominated by $P_{J|X,Y}$,
completing the proof of Theorem \ref{thm:main}.

\medskip{}

We also remark that the $K$ in Lemma \ref{lem:pfr} satisfies
\begin{equation}
\mathbb{E}[\log K]\le D_{\mathrm{KL}}(P\Vert Q)+1,\label{eq:new_logk}
\end{equation}
which slightly improves upon the $\mathbb{E}[\log K]\le D_{\mathrm{KL}}(P\Vert Q)+e^{-1}\log e+1$
in \cite{sfrl_trans} and $\mathbb{E}[\log K]\le D_{\mathrm{KL}}(P\Vert Q)+\log e$
in \cite{li2021unified}. To prove this, writing $f(u):=(\mathrm{d}P/\mathrm{d}Q)(u)$,
and letting $U'\sim Q$ be independent of $U$, we have
\begin{align}
\mathbb{E}[\log K] & \le\mathbb{E}[\log\mathbb{E}[K|U]]\nonumber \\
 & =\mathbb{E}[\log\mathbb{E}[\max\{f(U),\,f(U')\}]]\nonumber \\
 & \le\mathbb{E}[\log\mathbb{E}[f(U)+f(U')]]\nonumber \\
 & =\mathbb{E}[\log(f(U)+1)]\nonumber \\
 & =\mathbb{E}[\log f(U)]+\mathbb{E}\left[\log\left(1+1/f(U)\right)\right]\nonumber \\
 & \le\mathbb{E}[\log f(U)]+\log\left(1+\mathbb{E}\left[1/f(U)\right]\right)\nonumber \\
 & =D_{\mathrm{KL}}(P\Vert Q)+\log\Big(1+\int\frac{1}{(\mathrm{d}P/\mathrm{d}Q)(u)}\mathrm{d}P(u)\Big)\nonumber \\
 & \le D_{\mathrm{KL}}(P\Vert Q)+\log2.\label{eq:new_logk_pf}
\end{align}
This offers a slight improvement for the strong functional representation
lemma \cite{sfrl_trans} to the following statement: for any jointly
distributed random variables $X,Y$, there exists a random variable
$Z$ independent of $X$ such that $Y$ is a function of $(X,Z)$,
and 
\begin{equation}
H(Y|Z)\le I(X;Y)+\log(I(X;Y)+2)+2.\label{eq:new_sfrl}
\end{equation}
Using the same arguments as in \eqref{eq:new_logk_pf}, we can show
that the $J$ in Theorem \ref{thm:main} satisfies
\begin{equation}
\mathbb{E}[\log J]\le\mathbb{E}[D_{\mathrm{KL}}(P_{Y|X}(\cdot|X)\Vert Q_{Y})]+1.\label{eq:new_logj}
\end{equation}
Note that $\mathbb{E}[D_{\mathrm{KL}}(P_{Y|X}(\cdot|X)\Vert Q_{Y})]=I(X;Y)$
when $Q_{Y}=P_{Y}$.

\medskip{}

\section{Expected Length}

In the following sections, we will present various consequences of
Theorem \ref{thm:main}. For example, substituting $\psi_{1}(x,y,k)=d(x,y)$
to be the distortion function, and $\psi_{2}(x,y,k)=\log k$, we have
the following variable-length lossy source coding result similar to
\cite{sfrl_trans} (with slightly improved constants), showing that
we can achieve an expected length close to the rate-distortion function
$R(D)$ even in one-shot.

\medskip{}

\begin{cor}
[Variable-length lossy compression]Fix any $P_{X}$, distortion
function $d:\mathcal{X}\times\mathcal{Y}\to\mathbb{R}$ and $D\in\mathbb{R}$.
Then there exists a lossy compression scheme with description $K\in\mathbb{Z}_{>0}$
and reconstruction $\tilde{Y}$ such that $\mathbb{E}[d(X,\tilde{Y})]\le D$,
and
\[
\mathbb{E}[\log K]\le R(D)+2.01,
\]
where $R(D):=\inf_{P_{Y|X}:\,\mathbb{E}[d(X,Y)]\le D}I(X;Y)$ is the
rate-distortion function. Hence, there exists a lossy compression
scheme with description $M\in\{0,1\}^{*}$ without the prefix-free
condition, and reconstruction $\tilde{Y}$, such that $\mathbb{E}[d(X,\tilde{Y})]\le D$,
and
\[
\mathbb{E}[|M|]\le R(D)+2.01.
\]
If we require the prefix-free condition, we instead have
\begin{equation}
\mathbb{E}[|M|]\le R(D)+\log(R(D)+3.01)+4.01.\label{eq:em_vl}
\end{equation}
\end{cor}
\medskip{}

\begin{IEEEproof}
Consider any $P_{Y|X}$. Let $(X,Y)\sim P_{X}P_{Y|X}$ and $Q_{Y}=P_{Y}$.
Construct a scheme by applying Theorem \ref{thm:main} on $\psi_{1}(x,y,k)=d(x,y)$
and $\psi_{2}(x,y,k)=\log k$. We have $\mathbb{E}[d(X,\tilde{Y})]\le\mathbb{E}[d(X,Y)]$
and $\mathbb{E}[\log K]\le\mathbb{E}[\log(2J)]\le I(X;Y)+2$ by \eqref{eq:new_logj}.
By considering $P_{Y|X}$ approaching the rate-distortion function,
we can have $\mathbb{E}[\log K]\le R(D)+2.01$. The result for variable-length
non-prefix-free description is because $K$ can be encoded into $\lfloor\log K\rfloor\le\log K$
bits. The result for prefix-free description follows from the inequality
$H(K)\le\mathbb{E}[\log K]+\log(\mathbb{E}[\log K]+1)+1$ in \cite{sfrl_trans},
and the application of Huffman coding.
\end{IEEEproof}
\smallskip{}

We remark that it is possible to improve \eqref{eq:em_vl} to 
\begin{equation}
\mathbb{E}[|M|]\le R(D)+\log(R(D)+2)+4.01\label{eq:em_vl_improve}
\end{equation}
by applying the derandomization step in the proof of Theorem \ref{thm:main}
after we construct the variable-length code. By \eqref{eq:new_logk},
if the encoder and decoder are allowed to share the Poisson process,
we can have $\mathbb{E}[\log K]\le I(X;Y)+1$, which can be compressed
into a prefix-free description $M$ with $\mathbb{E}[|M|]\le I(X;Y)+\log(I(X;Y)+2)+3$
due to $H(K)\le\mathbb{E}[\log K]+\log(\mathbb{E}[\log K]+1)+1$ \cite{sfrl_trans}.
The extra $1$- bit penalty comes from a similar derandomization step
as in the proof of Theorem \ref{thm:main}, and the ``$0.01$''
penalty (which can be made arbitrarily small) is to accomodate for
the situation where the infimum in $R(D)$ cannot be attained.

\smallskip{}

\section{Notions of Pointwise Redundancy\label{sec:redundancy} }

The pointwise redundancy of a prefix-free lossless source code $f:\mathcal{X}\to\{0,1\}^{*}$
for the distribution $P_{X}$ at $X\in\mathcal{X}$ is $|M|+\log P_{X}(X)$
\cite{drmota2004precise}. The $|M|$ corresponds to the actual length
of the encoding $M=f(X)\in\{0,1\}^{*}$ of $X$, whereas $-\log P_{X}(X)$
is the ``amount of information'' in $X$. For lossy compression,
it is not entirely clear how to measure the ``amount of information''
in $(X,\tilde{Y})$ (where $\tilde{Y}$ is the actual reconstruction
of the compression). To allow full generality, in this paper, we define
the $\eta$\emph{-pointwise redundancy} (where $\eta:\mathcal{X}\times\mathcal{Y}\to\mathbb{R}$
is a function) as
\[
|M|-\eta(X,\tilde{Y}).
\]
Here $\eta(X,\tilde{Y})$ plays the role of the ``amount of information''
in $(X,\tilde{Y})$. We discuss several examples of $\eta$-pointwise
redundancy.

\smallskip{}

\begin{itemize}
\item \textbf{Pointwise rate redundancy (PRR)}, studied in \cite{kontoyiannis2000pointwise,ouguz2012pointwise},
is given by
\[
|M|-R(D),
\]
i.e., we take $\eta(x,y)=R(D)$ to be the rate-distortion function
at $D$, where $D=\mathbb{E}[d(X,\tilde{Y})]$ is the expected distortion
of the scheme. Unlike the pointwise redundancy for lossless compression,
here the ``amount of information'' term $\eta(x,y)$ does not depend
on the values of $x,y$. 
\end{itemize}
\smallskip{}

\begin{itemize}
\item \textbf{Pointwise source-wise redundancy} \textbf{(PSR)}, studied
in \cite{kontoyiannis2000pointwise},\footnote{The results on both PRR and PSR are called ``pointwise redundancy''
in \cite{kontoyiannis2000pointwise}. Here we use the names ``pointwise
rate redundancy'' and ``pointwise source-wise redundancy'' to distinguish
them.} is given by
\[
|M|-\jmath(X,D),
\]
i.e., we take $\eta(x,y)=\jmath(x,D)$, where $D=\mathbb{E}[d(X,\tilde{Y})]$,
and $\jmath(x,D)$ is the \emph{$d$-tilted information in }$x$ \cite{csiszar1974extremum,kontoyiannis2000pointwise,kostina2012fixed}
\begin{equation}
\jmath(x,D):=-\log\mathbb{E}\left[2^{-\lambda^{*}(d(x,Y^{*})-D)}\right],\label{eq:tilted}
\end{equation}
where $Y^{*}\sim P_{Y}$ follows the $Y$-marginal of $P_{X}P_{Y|X}$
where $P_{Y|X}$ is the conditional distribution that attains the
minimum in $R(D)$ (assume that the minimizer is unique), and $\lambda^{*}:=-R'(D)$
is the negative of the derivative of the rate-distortion function
at $D$. Note that $\eta(x,y)$ only depends on $x$. The $d$-tilted
information is considered as an analogue of the amount of information
$-\log P_{X}(X)$ in lossless source coding \cite{kostina2012fixed},
and hence the pointwise source redundancy can be considered as an
analogue of the pointwise redundancy in lossless source coding. In
\cite{kontoyiannis2000pointwise}, it has been shown that when the
source $X=X^{n}$ is an i.i.d. sequence, and we require $d(X^{n},\tilde{Y}^{n})=n^{-1}\sum_{i=1}^{n}d(X_{i},\tilde{Y}_{i})\le D$
almost sure (i.e., this is a $D$-semifaithful code), the every sequence
of codes (indexed by the blocklength $n$) must have a PSR
\[
|M|-\jmath(X^{n},D)\ge-2\log n
\]
eventually as $n\to\infty$ with probability $1$. Also, there exists
codes with PSR
\[
|M|-\jmath(X^{n},D)\le5\log n
\]
eventually as $n\to\infty$ with probability $1$.
\end{itemize}
\smallskip{}

\begin{itemize}
\item \textbf{Pointwise source-distortion-wise redundancy} \textbf{(PSDR)},
defined as 
\[
|M|-\jmath(X,D,d(X,\tilde{Y})),
\]
where we write
\begin{align}
\jmath(x,D,\delta) & :=-\log\mathbb{E}\left[2^{-\lambda^{*}(d(x,Y^{*})-\delta)}\right]\nonumber \\
 & =\jmath(x,D)-\lambda^{*}(\delta-D),\label{eq:tilted_dis}
\end{align}
where $Y^{*}$ and $\lambda^{*}$ are the same as \eqref{eq:tilted}.
Here $\eta(x,y)=\jmath(x,D,d(x,y))$ depends on both $x$ and the
distortion $\delta=d(x,y)$, and can be interpreted as ``the amount
of information required to convey $x$ within a distortion $\delta$''.
Invoking \cite[Lemma 1.4]{csiszar1974extremum} (also see \cite[Property 1]{kostina2012fixed}),
for $P_{Y}$-almost all $y$,
\begin{align}
\jmath(x,D,d(x,y)) & =\iota_{X;Y}(x;y),\label{eq:tilted_iota}
\end{align}
where $P_{Y}$ is the $Y$-marginal of $(X,Y)\sim P_{X}P_{Y|X}$,
and $P_{Y|X}$ attains the minimum in $R(D)$. Hence, when $\mathcal{Y}$
is finite, the PSDR equals $|M|-\iota_{X;Y}(X;\tilde{Y})$ if $\tilde{Y}$
is in the support of $P_{Y}$.
\end{itemize}
\smallskip{}

While the $\eta(x,y)$'s in PRR, PSR and PSDR all corresponds to ``the
amount of information required to convey the source within some distortion'',
their difference lies in their ``level of pointwise-ness''. The
$\eta$ in PRR is the ``global average amount of information'' that
does not depend on the point $(x,y)$. The $\eta$ in PSR is ``source-wise'',
in the sense that it is the ``average amount of information at $x$''
that only depends on the source $x$. The $\eta$ in PSDR is ``source-and-distortion-wise'',
in the sense that it is the ``amount of information at $x$ and distortion
$\delta$'' that depends on both $x$ and the distortion $\delta$
between the current source and reconstruction (not only the average
distortion). Also note that both PSR and PSDR (but not PRR) can recover
the pointwise redundancy $|M|+\log P_{X}(X)$ for lossless source
coding by taking $d(x,y)=\mathbf{1}\{x\neq y\}$.

All three choices of $\eta$'s have expectation $R(D)$. We have $\mathbb{E}[\jmath(X,D)]=R(D)$
as proved in \cite{csiszar1974extremum}. For PSDR, we have $\mathbb{E}[\jmath(X,D,d(X,\tilde{Y}))]=\mathbb{E}[\jmath(X,D)-\lambda^{*}(d(X,\tilde{Y})-D)]=R(D)$
since $D=\mathbb{E}[d(X,\tilde{Y})]$. Therefore, when $M$ is a prefix-free
codeword, the expectation of each of the three pointwise redundancies
must be nonnegative.

\smallskip{}

\section{Pointwise Redundancy without Prefix-free Condition\label{sec:noprefix}}

If we use a variable-length code without the prefix-free condition,
then a positive integer description $K\in\mathbb{Z}_{>0}$ can be
encoded into $\lfloor\log K\rfloor\le\log K$ bits. Hence, we can
bound the pointwise redundancy by $\log K-\eta(X;\tilde{Y})$. Without
the prefix-free condition, the expectation of the pointwise redundancies
may be negative, though the gap is at most logarithmic \cite{szpankowski2011minimum}.
The following corollary of Theorem \ref{thm:main} gives a bound for
the pointwise redundancy for general $\eta(x,y)$. 

\medskip{}

\begin{cor}
[Pointwise redundancy w/o prefix-free condition]Fix any $P_{X}$,
$P_{Y|X}$, distortion function $d:\mathcal{X}\times\mathcal{Y}\to[0,\infty)$,
and function $\eta:\mathcal{X}\times\mathcal{Y}\to\mathbb{R}$. Then
there exists a lossy compression scheme with description $K\in\mathbb{Z}_{>0}$
and reconstruction $\tilde{Y}$ such that $\mathbb{E}[d(X,\tilde{Y})]\le\mathbb{E}[d(X,Y)]$,
and
\begin{align}
 & \mathbb{P}\left(\log K-\eta(X,\tilde{Y})\ge\gamma\right)\nonumber \\
 & \le2^{-\gamma+1}\mathbb{E}\left[2^{-\eta(X,Y)}(2^{\iota_{X;Y}(X;Y)}+1)\right]\label{eq:point_wop}
\end{align}
for every $\gamma\in\mathbb{R}$, where $(X,Y)\sim P_{X}P_{Y|X}$. 
\end{cor}
\smallskip{}

The result specialized for PSDR is especially elegant.

\smallskip{}

\begin{cor}
[PSDR w/o prefix-free condition] \label{cor:psdr_noprefix}For $D>0$,
under the regularity conditions in \cite{kostina2012fixed},\footnote{The regularity conditions in \cite{kostina2012fixed} are: $R(\delta)$
is finite for some $\delta$, there exists a finite set $\mathcal{E}\subseteq\mathcal{Y}$
such that $\mathbb{E}[\min_{y\in\mathcal{E}}d(X,y)]<\infty$, and
the minimum in $R(D)$ is achieved by a unique $P_{Y|X}$.} there exists a lossy compression scheme with description $K\in\mathbb{Z}_{>0}$,
reconstruction $\tilde{Y}$, with $\mathbb{E}[d(X,\tilde{Y})]\le D$,
and with pointwise source-distortion redundancy (see \eqref{eq:tilted_dis})
satisfying
\[
\mathbb{P}\left(\log K-\jmath(X,D,d(X,\tilde{Y}))\ge\gamma\right)\le2^{-\gamma+2}
\]
for every $\gamma\in\mathbb{R}$. The above is also true when $\jmath(X,D,d(X,\tilde{Y}))$
is replaced with $\iota_{X;Y}(X;\tilde{Y})$, where $P_{Y|X}$ attains
the minimum in $R(D)$.
\end{cor}
\medskip{}

We now prove these two results.
\begin{IEEEproof}
Construct a scheme by applying Theorem \ref{thm:main} on $\psi_{1}(x,y,k)=d(x,y)$
and $\psi_{2}(x,y,k)=2^{-\eta(x,y)}k$. We have
\begin{align*}
 & \mathbb{P}\left(\log K-\eta(X,\tilde{Y})\ge\gamma\right)\\
 & =\mathbb{P}\left(2^{-\eta(X,\tilde{Y})-\gamma}K\ge1\right)\\
 & \le\mathbb{E}\left[2^{-\eta(X,\tilde{Y})-\gamma}K\right]\\
 & =2^{-\gamma}\mathbb{E}\left[2^{-\eta(X,\tilde{Y})}K\right]\\
 & \stackrel{(a)}{\le}2^{-\gamma}\mathbb{E}\left[2^{-\eta(X,Y)}2J\right]\\
 & =2^{-\gamma+1}\mathbb{E}\left[2^{-\eta(X,Y)}\mathbb{E}[J\,|\,X,Y]\right]\\
 & \stackrel{(b)}{=}2^{-\gamma+1}\mathbb{E}\left[2^{-\eta(X,Y)}(2^{\iota_{X;Y}(X;Y)}+1)\right],
\end{align*}
where (a) is by \eqref{eq:main_exp}, and (b) is by \eqref{eq:main_geom}.
We now consider PSDR where $\eta(x,y)=\jmath(x,D,d(x,y))$. Consider
$P_{Y|X}$ attaining the minimum in $R(D)$. We have
\begin{align*}
 & \mathbb{P}\left(\log K-\jmath(X,D,d(X,\tilde{Y}))\ge\gamma\right)\\
 & \le2^{-\gamma+1}\mathbb{E}\left[2^{-\jmath(X,D,d(X,Y))}(2^{\iota_{X;Y}(X;Y)}+1)\right]\\
 & \stackrel{(c)}{=}2^{-\gamma+1}\mathbb{E}\left[2^{-\iota_{X;Y}(X;Y)}(2^{\iota_{X;Y}(X;Y)}+1)\right]\\
 & =2^{-\gamma+1}\left(\mathbb{E}\left[2^{-\iota_{X;Y}(X;Y)}\right]+1\right)\\
 & \stackrel{(d)}{\le}2^{-\gamma+2},
\end{align*}
where (c) is by \eqref{eq:tilted_iota}, and (d) is because $\mathbb{E}[2^{-\iota_{X;Y}(X;Y)}]=\int((\mathrm{d}P_{X,Y}/\mathrm{d}P_{X}P_{Y})(x,y))^{-1}P_{X,Y}(\mathrm{d}x,\mathrm{d}y)\le1$.
\end{IEEEproof}
\medskip{}

The term inside the expectation in \eqref{eq:point_wop} is unbounded,
which might make the expectation problematic to bound, for example,
for PRR and PSR. Alternatively, we can also have the following result
with a bounded expectation, which gives a meaningful error bound for
PRR and PSR. The downside is that the scheme has to be designed for
a specific $\gamma$.

\medskip{}

\begin{cor}
[Pointwise redundancy w/o prefix-free condition]Fix any $P_{X}$,
$P_{Y|X}$, distortion function $d:\mathcal{X}\times\mathcal{Y}\to[0,\infty)$,
function $\eta:\mathcal{X}\times\mathcal{Y}\to\mathbb{R}$ and $\gamma\in\mathbb{R}$.
Then there exists a lossy compression scheme with description $K\in\mathbb{Z}_{>0}$
and reconstruction $\tilde{Y}$ such that $\mathbb{E}[d(X,\tilde{Y})]\le\mathbb{E}[d(X,Y)]$,
and
\begin{align*}
 & \mathbb{P}\left(\log K-\eta(X,\tilde{Y})\ge\gamma\right)\\
 & \le\mathbb{E}\left[\min\big\{2^{-\eta(X,Y)-\gamma+1}(2^{\iota_{X;Y}(X;Y)}+1),\,1\big\}\right],
\end{align*}
where $(X,Y)\sim P_{X}P_{Y|X}$. 
\end{cor}
\medskip{}

\begin{IEEEproof}
Construct a scheme by applying Theorem \ref{thm:main} on $\psi_{1}(x,y,k)=d(x,y)$
and $\psi_{2}(x,y,k)=\min\{2^{-\eta(x,y)-\gamma}k,\,1\}$. We have
\begin{align*}
 & \mathbb{P}\left(\log K-\eta(X,\tilde{Y})\ge\gamma\right)\\
 & \le\mathbb{E}\left[\min\big\{2^{-\eta(X,\tilde{Y})-\gamma}K,\,1\big\}\right]\\
 & \le\mathbb{E}\left[\min\big\{2^{-\eta(X,Y)-\gamma+1}J,\,1\big\}\right]\\
 & \le\mathbb{E}\left[\min\big\{2^{-\eta(X,Y)-\gamma+1}\mathbb{E}[J\,|\,X,Y],\,1\big\}\right]\\
 & \le\mathbb{E}\left[\min\big\{2^{-\eta(X,Y)-\gamma+1}(2^{\iota_{X;Y}(X;Y)}+1),\,1\big\}\right].
\end{align*}
\end{IEEEproof}
\smallskip{}

\section{Pointwise Redundancy with Prefix-free Condition\label{sec:prefix}}

If a prefix-free coding scheme is desired, we can apply the Elias
delta code \cite{elias1975universal} to obtain the following slightly
more complicated result.

\medskip{}

\begin{cor}
[Pointwise redundancy with prefix-free condition]Fix any $P_{X}$,
$P_{Y|X}$, distortion function $d:\mathcal{X}\times\mathcal{Y}\to[0,\infty)$,
function $\eta:\mathcal{X}\times\mathcal{Y}\to\mathbb{R}$, and $\gamma\in\mathbb{R}$.
Then there exists a lossy compression scheme with prefix-free description
$M\in\{0,1\}^{*}$ and reconstruction $\tilde{Y}$ such that $\mathbb{E}[d(X,\tilde{Y})]\le\mathbb{E}[d(X,Y)]$,
and
\begin{align*}
 & \mathbb{P}\left(|M|-\eta(X,\tilde{Y})\ge\gamma\right)\\
 & \le\mathbb{E}\Big[\min\big\{2^{-\eta(X,Y)-\gamma+2}([\eta(X,Y)+\gamma]_{+}+1)^{2}\\
 & \;\;\;\;\;\;\;\;\;\cdot(2^{\iota_{X;Y}(X;Y)}+1),\,1\big\}\Big],
\end{align*}
where $(X,Y)\sim P_{X}P_{Y|X}$. 
\end{cor}
\smallskip{}

The result specialized for PSDR is given below.

\smallskip{}

\begin{cor}
[PSDR with prefix-free condition] For $D>0$, $\gamma\in\mathbb{R}$,
under the regularity conditions in \cite{kostina2012fixed} (see Corollary
\ref{cor:psdr_noprefix}), there exists a lossy compression scheme
with prefix-free description $M\in\{0,1\}^{*}$, reconstruction $\tilde{Y}$,
with $\mathbb{E}[d(X,\tilde{Y})]\le D$, and with pointwise source-distortion
redundancy (see \eqref{eq:tilted_dis}) satisfying
\begin{align*}
 & \mathbb{P}\left(|M|-\jmath(X,D,d(X,\tilde{Y}))\ge\gamma\right)\\
 & \le2^{-\gamma+3}\mathbb{E}\big[([\iota_{X;Y}(X;Y)+\gamma]_{+}+1)^{2}\big],
\end{align*}
where $(X,Y)\sim P_{X}P_{Y|X}$, and $P_{Y|X}$ attains the minimum
in $R(D)$. 
\end{cor}
\medskip{}

We now prove the two results.
\begin{IEEEproof}
Let $L(t):=t+2\log(t+1)+1$. Let $L^{-1}$ be the inverse function
of $L$ (take $L^{-1}(a)=0$ for $a<1$). Construct a scheme by applying
Theorem \ref{thm:main} on $\psi_{1}(x,y,k)=d(x,y)$ and $\psi_{2}(x,y,k)=\min\{2^{-L^{-1}(\eta(x,y)+\gamma)}k,\,1\}$,
and encode the description $K$ by Elias delta code \cite{elias1975universal}
that takes $\le L(\log K)$ bits. We have
\begin{align*}
 & \mathbb{P}\left(|M|-\eta(X,\tilde{Y})\ge\gamma\right)\\
 & \le\mathbb{P}\left(L(\log K)-\eta(X,\tilde{Y})\ge\gamma\right)\\
 & \le\mathbb{P}\left(\log K\ge L^{-1}(\eta(X,\tilde{Y})+\gamma)\right)\\
 & =\mathbb{P}\left(2^{-L^{-1}(\eta(X,\tilde{Y})+\gamma)}K\ge1\right)\\
 & \le\mathbb{E}\left[\min\big\{2^{-L^{-1}(\eta(X,\tilde{Y})+\gamma)}K,\,1\big\}\right]\\
 & \stackrel{(a)}{\le}\mathbb{E}\left[\min\big\{2^{-L^{-1}(\eta(X,Y)+\gamma)+1}J,\,1\big\}\right]\\
 & \le\mathbb{E}\left[\min\big\{2^{-L^{-1}(\eta(X,Y)+\gamma)+1}\mathbb{E}[J\,|\,X,Y],\,1\big\}\right]\\
 & \stackrel{(b)}{=}\mathbb{E}\left[\min\big\{2^{-L^{-1}(\eta(X,Y)+\gamma)+1}(2^{\iota_{X;Y}(X;Y)}+1),\,1\big\}\right],
\end{align*}
where (a) is by \eqref{eq:main_exp}, and (b) is by \eqref{eq:main_geom}.
To bound $L^{-1}(a)$, note that for $a\in\mathbb{R}$ with $a-2\log([a]_{+}+1)-1\ge0$,
\begin{align*}
 & L(a-2\log([a]_{+}+1)-1)\\
 & =a-2\log(a+1)+2\log(a-2\log(a+1))\\
 & \le a,
\end{align*}
and $L^{-1}(a)\ge a-2\log([a]_{+}+1)-1$. This inequality holds for
all $a\in\mathbb{R}$ as well since $L^{-1}(a)\ge0$. Hence,
\begin{align*}
 & \mathbb{P}\left(|M|-\eta(X,\tilde{Y})\ge\gamma\right)\\
 & \le\mathbb{E}\left[\min\big\{2^{-\eta(X,Y)-\gamma+2\log([\eta(X,Y)+\gamma]_{+}+1)+2}(2^{\iota_{X;Y}(X;Y)}+1),\,1\big\}\right]\\
 & =\mathbb{E}\Big[\min\big\{2^{-\eta(X,Y)-\gamma+2}([\eta(X,Y)+\gamma]_{+}+1)^{2}(2^{\iota_{X;Y}(X;Y)}+1),\,1\big\}\Big].
\end{align*}
We now consider PSDR where $\eta(x,y)=\jmath(x,D,d(x,y))$. Consider
$P_{Y|X}$ attaining the minimum in $R(D)$. We have
\begin{align*}
 & \mathbb{P}\left(|M|-\iota_{X;Y}(X;\tilde{Y})\ge\gamma\right)\\
 & \le\mathbb{E}\Big[2^{-\eta(X,Y)-\gamma+2}([\eta(X,Y)+\gamma]_{+}+1)^{2}(2^{\iota_{X;Y}(X;Y)}+1)\Big].\\
 & \stackrel{(c)}{=}\mathbb{E}\Big[2^{-\gamma+2}([\iota_{X;Y}(X;Y)+\gamma]_{+}+1)^{2}(2^{-\iota_{X;Y}(X;Y)}+1)\Big]\\
 & \stackrel{(d)}{\le}2^{-\gamma+2}\mathbb{E}\big[([\iota_{X;Y}(X;Y)+\gamma]_{+}+1)^{2}\big]\mathbb{E}\left[2^{-\iota_{X;Y}(X;Y)}+1\right]\\
 & =2^{-\gamma+3}\mathbb{E}\big[([\iota_{X;Y}(X;Y)+\gamma]_{+}+1)^{2}\big],
\end{align*}
where (c) is by \eqref{eq:tilted_iota}, and (d) is by rearrangement
inequality.
\end{IEEEproof}

\medskip{}

\section{Lossy Gray-Wyner System}

We now generalize Theorem \ref{thm:main} to the lossy Gray-Wyner
system \cite{gray1974source}, which is a network with one encoder
and two decoders. In the one-shot lossy Gray-Wyner system (see \cite{sfrl_trans}),
the encoder observes the source pair $(X_{1},X_{2})\sim P_{X_{1},X_{2}}$,
and produces three descriptions $K_{0},K_{1},K_{2}\in\mathbb{Z}_{>0}$.
Decoder 1 observes $K_{0},K_{1}$, and outputs $\tilde{Y}_{1}$. Decoder
2 observes $K_{0},K_{2}$, and outputs $\tilde{Y}_{2}$. We are interested
in bounding the distortions $d_{1}(X_{1},\tilde{Y}_{1})$ and $d_{2}(X_{2},\tilde{Y}_{2})$,
where $d_{i}:\mathcal{X}_{i}\times\mathcal{Y}_{i}\to\mathbb{R}$ are
distortion functions. The following is an extension of Theorem \ref{thm:main}
to the one-shot lossy Gray-Wyner system.

\medskip{}

\begin{thm}
\label{thm:gw}Fix any $P_{X_{1},X_{2}}$, $P_{U|X_{1},X_{2}}$, $P_{Y_{1}|X_{1},U}$,
$P_{Y_{2}|X_{2},U}$. Consider any collection of functions $\psi_{i}:\mathcal{X}_{1}\times\mathcal{X}_{2}\times\mathcal{Y}_{1}\times\mathcal{Y}_{2}\times\mathbb{Z}_{>0}^{3}\to\mathbb{R}$
that is nondecreasing in each of the last three arguments for $i=1,\ldots,\ell$.
Then there exists a lossy compression scheme for the one-shot lossy
Gray-Wyner system with positive integer descriptions $K_{0},K_{1},K_{2}\in\mathbb{Z}_{>0}$
and reconstructions $\tilde{Y}_{1},\tilde{Y}_{2}$ such that
\begin{align*}
 & \mathbb{E}\big[\psi_{i}(X_{1},X_{2},\tilde{Y}_{1},\tilde{Y}_{2},K_{0},K_{1},K_{2})\big]\\
 & \le\mathbb{E}\big[\psi_{i}(X_{1},X_{2},Y_{1},Y_{2},\ell J_{0},J_{1},J_{2})\big]
\end{align*}
for $i=1,\ldots,\ell$, where $(X_{1},X_{2},U,Y_{1},Y_{2})\sim P_{X_{1},X_{2}}P_{U|X_{1},X_{2}}P_{Y_{1}|X_{1},U}P_{Y_{2}|X_{2},U}$,
and $J_{0},J_{1},J_{2}\in\mathbb{Z}_{>0}$ are conditionally independent
conditional on $(X_{1},X_{2},U,Y_{1},Y_{2})$, and have the following
conditional distributions
\begin{align*}
 & J_{0}\,|\,(X_{1},X_{2},U,Y_{1},Y_{2})\\
 & \sim\mathrm{Geom}\big((2^{\iota_{U;X_{1},X_{2}}(U;X_{1},X_{2})}+1)^{-1}\big),
\end{align*}
and for $i=1,2$,
\begin{align*}
 & J_{i}\,|\,(X_{1},X_{2},U,Y_{1},Y_{2})\\
 & \sim\mathrm{Geom}\big((2^{\iota_{Y_{i};X_{i}|U}(Y_{i};X_{i}|U)}+1)^{-1}\big).
\end{align*}
\end{thm}
\medskip{}

\begin{IEEEproof}
Here we use a construction mostly similar to \cite[Theorem 4]{sfrl_trans}.
Generate marked Poisson processes $(\bar{U}_{i},T_{0,i})_{i}$, $(\bar{Y}_{1,i},T_{1,i})_{i}$
and $(\bar{Y}_{2,i},T_{2,i})_{i}$, where $\bar{U}_{i}\stackrel{iid}{\sim}P_{U}$,
$\bar{Y}_{1,i}\stackrel{iid}{\sim}P_{Y_{1}}$ and $\bar{Y}_{2,i}\stackrel{iid}{\sim}P_{Y_{2}}$
(see the proof of Theorem \ref{thm:main}). Given $X_{1},X_{2}$,
the encoder outputs 
\begin{equation}
K_{0}:=\mathrm{argmin}_{i}T_{0,i}2^{-\iota_{U;X_{1},X_{2}}(\bar{U}_{i};X_{1},X_{2})}.\label{eq:k_scheme-1}
\end{equation}
Let $U=\bar{U}_{K_{0}}$. Consider the process $(\bar{Y}_{1,i},T_{1,i}2^{-\iota_{U;Y_{1}}(U;\bar{Y}_{1,i})})_{i}$,
and let $(\bar{Y}'_{1,i},T'_{1,i})_{i}$ be the same process but with
the ``$T$'' coordinate sorted in ascending order, i.e., $T'_{1,1}\le T'_{1,2}\le\cdots$.
Note that $(\bar{Y}'_{1,i},T'_{1,i})_{i}$ is conditionally a marked
Poisson processes with $\bar{Y}'_{1,i}$ conditionally i.i.d. following
$P_{Y_{1}|U}$ given $U$ (see \cite[Definition 2]{li2021unified}).
The encoder produces
\[
K_{1}:=\mathrm{argmin}_{i}T'_{1,i}2^{-\iota_{Y_{1};X_{1}|U}(\bar{Y}'_{1,i};X_{1}|U)}.
\]
Define $(\bar{Y}'_{2,i},T'_{2,i})_{i}$ and $K_{2}$ similarly. Given
$K_{0},K_{1},K_{2}$, the decoder computes $U=\bar{U}_{K_{0}}$, and
outputs $Y_{1}=\bar{Y}'_{1,K_{1}}$, $Y_{2}=\bar{Y}'_{2,K_{1}}$.
By the same arguments as in the proof of Theorem \ref{thm:main},
$K_{0},K_{1},K_{2}$ are first order stochastically dominated by $J_{0},J_{1},J_{2}$
respectively. The proof is completed by invoking the same ``derandomization''
argument as in the proof of Theorem \ref{thm:main}, where we use
$K'_{0}=\ell(K_{0}-1)+j$ to communicate which codebook to use.
\end{IEEEproof}
\medskip{}

Note that $\mathbb{E}[\log J_{0}]\le I(U;X_{1},X_{2})+1$ and $\mathbb{E}[\log J_{i}]\le I(Y_{i};X_{i}|U)+1$
by \ref{eq:new_logk}. Therefore, Theorem \ref{thm:gw} implies the
asymptotic lossy Gray-Wyner rate region \cite{elgamal2011network}.
Nevertheless, it is unclear what is the correct notion of ``pointwise
redundancy'' in a setting with three descriptions. We leave the generalization
of the pointwise redundancy bounds in Sections \ref{sec:redundancy},
\ref{sec:noprefix}, \ref{sec:prefix} to the lossy Gray-Wyner system,
and possible connections to the second-order region \cite{watanabe2016second},
for future studies.

\medskip{}

\section{Acknowledgement}

The work described in this paper was partially supported by an ECS
grant from the Research Grants Council of the Hong Kong Special Administrative
Region, China {[}Project No.: CUHK 24205621{]}.

\medskip{}

\bibliographystyle{IEEEtran}
\bibliography{ref}

\end{document}